\documentclass[aps,pra,twocolumn,amsmath,showpacs]{revtex4}
\usepackage{bm}
\usepackage{graphicx}
\begin{document}
\setlength{\arraycolsep}{2pt}
\title{Demonstrating higher-order nonclassical effects by photon-added classical states: Realistic schemes}
\author{Juhui Lee$^{1,2}$, Jaewan Kim$^{1}$, and Hyunchul Nha$^{3,*}$}
\affiliation{$^1$School of Computational Sciences, Korea Institute for Advanced Study, Seoul, Korea}  
\affiliation{$^2$Department of Physics, Sookmyung Women's University, Seoul, Korea}  
\affiliation{$^3$Department of Physics, Texas A \& M University at Qatar, PO Box 23874, Doha, Qatar}
\date{\today}
%\maketitle
\begin{abstract}
Detecting nonclassical properties that do not allow classical interpretation of photoelectric counting events is 
one of the crucial themes in quantum optics. 
Observation of individual nonclassical effects for a single-mode field, however, has been so far practically confined  
to sub-Poissonian statistics and quadrature squeezing. 
We show that a photon-added classical (coherent or thermal) state exhibits generalized nonclassical features in all orders of creation and annihilation operators, 
thereby becoming a promising candidate for studying higher-order nonclassical effects. 
Our analysis demonstrates robustness of these effects against nonideal experimental conditions.
\end{abstract}
\maketitle
\email{hyunchul.nha@qatar.tamu.edu}

\narrowtext

\section{Introduction} 
%Nonclassical properties that cannot be explained by any classical stochastic theories are of central importance in quantum physics.
Ever since Glauber developed the concept of quantum optical coherence [1], 
nonclassical properties that cannot be explained by any classical, stochastic, theories 
have long been a subject of substantial theoretical and experimental efforts.
Nonclassicality in quantum optics is usually formulated in terms of the Glauber-Sudarshan $P$-representation [2,3], 
in which a single-mode state $\rho$ can be expressed as $\rho=\int d^2\alpha P(\alpha)|\alpha\rangle\langle\alpha|$ 
in the basis of coherent states $|\alpha\rangle$. 
If the quasi-distribution $P(\alpha)$ cannot be admitted as a classical probability density, 
the state is called nonclassical [4].  
However, as the $P$-function is generally not directly measurable in itself, 
there have been two different approaches to the identification of nonclassical states. 
One is to construct experimentally accessible distributions, 
most notably the Wigner function by homodyne detection [5]. 
The negative value in Wigner distribution, for example, becomes a sufficient, though not necessary, signature of nonclassicality. 
The other is to observe individual nonclassical effects such as antibunching [6,7], 
sub-Poissonian statistics [8], and squeezing [9] that do not allow 
classical interpretation of photoelectric counting events [10]. 
%In fact, the Glauber-Sudarshan $P$-function is closely related to the optical detection 
%that is usually described as the process of absorbing (annihilating) photons \cite{Glauber2}. 
%More precisely, experimentally measured correlation functions are intrinsically expressed in normal ordering of annihilation and creation operators, 
%which makes the $P$-representation so special in quantum optics. 

In view of the second approach, there have also been numerous proposals for detecting nonclassical effects 
in higher-orders beyond those well-established sub-Poissonian statistics and quadrature squeezing. 
To name a few, higher-order quadrature squeezing was suggested by Hong and Mandel [11], 
amplitude-squared squeezing by Hillery [12], and higher-order photon statistics by Lee [13], 
and by Agarwal and Tara [14]. 
Recently, Shchukin {\it et al.} derived a hierarchy of sufficient and necessary conditions 
for nonclassicality in terms of normally-ordered moments [15]. 
Despite the theoretical efforts, however, 
experimental verifications were mostly limited to sub-Poissonian and squeezing phenomena. 
Although it suffices to detect any of possible nonclassical properties in some cases, 
a richer class of high-order nonclassical effects may reveal the distinct nature of quantum phenomena more deeply and 
find new applications, e.g., in quantum information science. 
In particular, it has been proved that any single-mode nonclassical state becomes a sufficient resource 
to generate a two-mode entanglement via a beam-splitter [16]. 
Furthermore, Nha and Zubairy derived entanglement criteria by which such output entangled states can be detected 
in the light of single-mode nonclassical properties of input field [17]. 

In this paper, we consider two generalized classes of nonclassical effects. 
The first, phase-sensitive, class is formulated as
$\langle:(\Delta (a^me^{-i\phi}+a^{\dag m}e^{i\phi}))^2:\rangle <0$, 
where :: denotes normal ordering of the creation and the annihilation operators ($m$: integer) [18].
Note that $m=1$ case refers to the usual quadrature squeezing and $m=2$ to the amplitude-squared squeezing by Hillery [12]. 
The second, phase-insensitive, class is represented by the condition $\langle:(\Delta a^{\dag m}a^m)^2:\rangle <0$, 
where $m=1$ is the case of sub-Poissonian statistics. 
Note that due to the entanglement inequalities established in Ref. [17], the single-mode nonclassical properties studied here are one-to-one related to verifiable two-mode entanglements via beam splitter. 

We investigate photon-added classical (coherent and thermal) states   
to observe the above nonclassical effects in higher-orders. 
The $n$-photon-added scheme, denoted by the mapping $\rho\rightarrow a^{\dag n}\rho a^n$, 
was first proposed by Agarwal and Tara [19]. 
The nonclassical properties of the photon-added coherent or thermal states have been studied 
by many [14,19-22], and very recently demonstrated in experiment [23-26]. 
However, those studies were mostly confined to low-order nonclassical properties, or the negativity of Wigner function. 
Recently, Duc and Noh studied high-order nonclassical properties of 
{\it pure} photon-added coherent states [27]. 
Here, we want to ask whether one can demonstrate the higher-order nonclassical effects in realistic conditions. 
We therefore focus on {\it single}-photon added coherent and thermal states in two practical schemes based on beam-splitter (BS) and nondegenerate parametric amplifier (NDPA), respectively. 
We show that the generalized nonclassical effects are robust against experimental imperfections, which renders photon added classical states promising for applications involving higher-order nonclassical properties. 
Specifically, it is shown that the experimental results (NDPA scheme) 
reported by Zavatta {\it et al.} already imply those high-order nonclassical effects.

\section{Generalized nonclassical properties}
Let us start by addressing a general nonclassical property 
for a single-mode state $\rho=\int d^2\alpha P(\alpha)|\alpha\rangle\langle\alpha|$. 
If the state $\rho$ has a classical probability density $P(\alpha)$, 
all normally-ordered positive operators must yield nonnegative ensemble averages. 
More precisely, for every operator $\hat f=\hat f(a^\dag,a)$ expressed in terms of the creation and the annihilation operator, $a^\dag$ and $a$,   
the quantum average $\langle :{\hat f}^\dag {\hat f}:\rangle$ becomes positive, 
$\langle :{\hat f}^\dag {\hat f}:\rangle=\int d^2\alpha |f(\alpha^*,\alpha)|^2P(\alpha)\ge 0$, 
if $P(\alpha)$ behaves like a probability density. 
In other words, $\langle :{\hat f}^\dag {\hat f}:\rangle<0$ is a clear signature of nonclassicality. 
In this paper, we take two generalized classes of operators ${\hat f}$ to construct higher-order nonclassical effects. 

{\bf (i) higher-order amplitude squeezing}\\
First, take ${\hat f}_1=\Delta (a^me^{-i\phi}+a^{\dag m}e^{i\phi})$, 
where $\Delta {\hat O}\equiv {\hat O}-\langle{\hat O}\rangle$ represents a quantum fluctuation and $\phi$ a phase angle ($m$: integer). 
Then, if there exists an angle $\phi$ such that
\begin{eqnarray}
\langle :{\hat f}_1^\dag {\hat f}_1:\rangle=\langle:(\Delta (a^me^{-i\phi}+a^{\dag m}e^{i\phi}))^2:\rangle <0,
\end{eqnarray} 
the state under consideration is nonclassical [18]. 
Note that $m=1$ is the case of quadrature squeezing, 
and $m=2$ that of amplitude-squared squeezing due to Hillery [12]. 
To define a quantity representing the depth of nonclassicality in the range of [-1,0), we note  
\begin{eqnarray}
&&\langle(\Delta (a^me^{-i\phi}+a^{\dag m}e^{i\phi}))^2\rangle\nonumber\\
&&=\langle:(\Delta (a^me^{-i\phi}+a^{\dag m}e^{i\phi}))^2:\rangle+
\langle [a^m,a^{\dag m}]\rangle.
\end{eqnarray} 
Therefore, we define a nonclassical depth 
\begin{eqnarray}
Q_1^m(\phi)=\frac{\langle:(\Delta (a^me^{-i\phi}+a^{\dag m}e^{i\phi}))^2:\rangle}
{\langle a^ma^{\dag m}\rangle-\langle a^{\dag m}a^m\rangle}, 
\label{eqn:NCD1}
\end{eqnarray}
which certainly takes a negative value only in the range [-1,0) for a nonclassical state. 
Note that the denominator of Eq.~(\ref{eqn:NCD1}) is always nonnegative. [See also Eq.~(\ref{eqn:ANM})].
%The negativity of $Q_1^m(\phi)$ in Eq.~(\ref{eqn:NCD1}) for a certain phase $\phi$ is a signature of nonclassicality. 

Now, instead of investigating the dependence of $Q_1^m(\phi)$ on the phase angle $\phi$, we can try to minimize $Q_1^m(\phi)$ over $\phi$. 
On expanding the terms, we find that 
\begin{eqnarray}
&&\langle:(\Delta (a^me^{-i\phi}+a^{\dag m}e^{i\phi}))^2:\rangle\nonumber\\
&&=\zeta e^{-2i\phi}+\zeta^*e^{2i\phi}+2\langle a^{\dag m}a^{m}\rangle-2\langle a^{\dag m}\rangle\langle a^{m}\rangle,
\end{eqnarray}
where
\begin{eqnarray}
\zeta\equiv\langle a^{\dag 2m}\rangle-\langle a^{\dag m}\rangle^2.
\end{eqnarray}
Using the fact that $\zeta e^{-2i\phi}+\zeta^*e^{2i\phi}$ takes the minimum value $-2|\zeta|$, 
the optimized nonclassical depth over the phases turns out to be
\begin{eqnarray}
Q_1^m=\frac{-2|\langle a^{\dag 2m}\rangle-\langle a^{\dag m}\rangle^2|
+2\langle a^{\dag m}a^{m}\rangle-2|\langle a^{\dag m}\rangle|^2}
{\langle a^ma^{\dag m}\rangle-\langle a^{\dag m}a^m\rangle},
\label{eqn:ONCD1}
\end{eqnarray}
which will be considered in this paper.

{\bf (ii) higher-order coincidence statistics}\\
Second, take ${\hat f}_2=\Delta (a^{\dag m}a^m)$, then we obtain another nonclassicality condition
\begin{eqnarray}
\langle :{\hat f}_2^\dag {\hat f}_2:\rangle=&&\langle:(\Delta (a^{\dag m}a^m)^2:\rangle \nonumber\\
=&&\langle a^{\dag 2m}a^{2m}\rangle-\langle a^{\dag m}a^{m}\rangle^2<0.
\end{eqnarray} 
This class includes the case of sub-Poissonian statistics for $m=1$. 
We define the second nonclassical depth 
\begin{eqnarray}
Q_2^m=\frac{\langle a^{\dag 2m}a^{2m}\rangle}
{\langle a^{\dag m}a^{m}\rangle^2}-1, 
\label{eqn:NCD2}
\end{eqnarray}
where $Q_2^{m=1}=\frac{\langle a^{\dag 2}a^{2}\rangle}
{\langle a^{\dag}a\rangle^2}-1$ is related to the Mandel-$Q$ factor as $Q_2^{m=1}=\frac{Q}{\langle a^{\dag}a\rangle}$. 
We note that the quantity $Q_2^m$ defined as such is, desirably, insensitive 
to the quantum efficiency of photo detectors in multiple-coincidence counting experiment, 
 which will be further addressed later.

\section{Single-photon added classical states}
In this section, we investigate the nonclassical properties of {\it pure} single-photon added coherent and thermal states before dealing with practical schemes in the next section. 
\subsection{single-photon added coherent state}
A single-photon added coherent state (SACS) is represented by $|\Psi\rangle=\frac{1}{\sqrt{1+\alpha^2}}a^\dag|\alpha\rangle$, 
where $\alpha$ is taken as a real parameter for simplicity, but without loss of generality. 
Using the relation $a^na^\dag=a^\dag a^n+na^{n-1}$ and its extensions, it is straightforward to obtain
\begin{eqnarray}
\langle a^{m}\rangle_{\rm SACS}&=&\frac{\alpha^m}{1+\alpha^2}(m+1+\alpha^2)\nonumber\\
\langle a^{\dag m}a^{m}\rangle_{\rm SACS}&=&\frac{\alpha^{2m+2}+(2m+1)\alpha^{2m}+m^2\alpha^{2m-2}}{1+\alpha^2}.\nonumber\\
\end{eqnarray}
Then, the numerator of $Q_1^m$ in Eq.~(\ref{eqn:ONCD1}) turns out to be
\begin{eqnarray}
&&-|\langle a^{\dag 2m}\rangle-\langle a^{\dag m}\rangle^2|
+\langle a^{\dag m}a^{m}\rangle-|\langle a^{\dag m}\rangle|^2\nonumber\\
&&=-\frac{m^2\alpha^{2m-2}}{(1+\alpha^2)^2}(\alpha^2-1),
\end{eqnarray}
which is negative for $\alpha>1$ in any order $m$. 
Therefore, starting with the coherent state with amplitude larger than unity, 
the resulting single-photon added state exhibits the generalized nonclassical effects in all orders.  
On the other hand, $Q_2^m$ turns out, for $\alpha\ne0$, to be 
\begin{eqnarray}
Q_2^m=\alpha^2(1+\alpha^2)\frac{\alpha^4+(4m+1)\alpha^2+4m^2}{(\alpha^4+(2m+1)\alpha^2+m^2)^2}-1<0,\nonumber\\
\end{eqnarray}
which is negative regardless of $\alpha$ and $m$. 
Thus, the second class of nonclassical effects appears for any single-photon added coherent states except for $\alpha=0$. 
In fact, the case of $\alpha=0$ corresponds to none other than the single photon Fock state, $a^\dag|0\rangle=|1\rangle$, 
which shows the nonclassical effect only in the lowest order of $m=1$, i.e., sub-Poissonian statistics. 

\subsection{single-photon added thermal state}
A single-photon added thermal state (SATS) is represented by $\rho=\frac{a^\dag\rho_{\rm th}a}{{\rm Tr}[a^\dag\rho_{\rm th}a]}$, 
where $\rho_{\rm th}=(1-e^{-\beta})e^{-\beta a^\dag a}$ is a thermal state with the average photon number $\bar{n}=\frac{1}{e^\beta-1}$. 
The SATS does not possess any phase-sensitive properties as it is symmetric in the phase-space distribution [25], 
which excludes the first class of nonclassical properties, $Q_1^m$. 
On the other hand, we obtain 
\begin{eqnarray}
\langle a^{\dag m}a^{m}\rangle_{\rm SATS}=\frac{m!}{x^m}(1+m(1+x)),
\end{eqnarray}
where $x\equiv e^\beta=1+\bar{n}^{-1}$. 
Therefore, $Q_2^m$ becomes negative
\begin{eqnarray}
Q_2^m=\frac{(2m)!}{(m!)^2}\frac{2m(1+x)+1}{(m(1+x)+1)^2}-1<0,
\end{eqnarray}
for 
\begin{eqnarray}
x>C_m\equiv\frac{(2m)!-(m!)^2+\sqrt{(2m)!((2m)!-(m!)^2)}}{m(m!)^2}.\nonumber\\
\end{eqnarray}
Note that $C_m$ is monotonically increasing with $m$, therefore, high-order nonclassical effects appear in the region of lower thermal-photon number.
This nonclassicality will be analyzed in more detail with experimental imperfections included in the next section.

\section{Realistic schemes}
In this section, we consider two realistic schemes to implement single-photon addition and analyze the nonclassical properties of the generated states. 
The first is the scheme proposed by Dakna {\it et al.}, where a classical state, coherent or thermal, and a single-photon source are injected to a beam-splitter to generate the single-photon added state conditioned on the non-detection of photons at one output [20,21]. 
The other is the one using the nondegenerate parametric amplifier that was already realized in Ref. [23-26]. 

\begin{figure}
\includegraphics[width=3.0in,keepaspectratio=true]{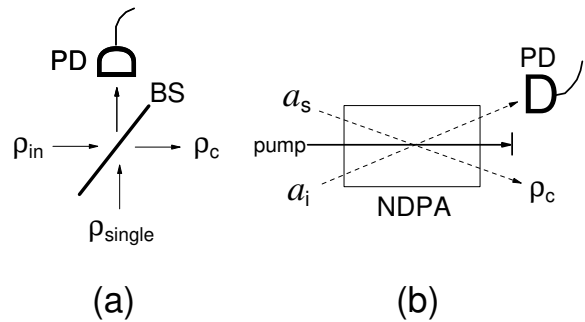}
\caption{Experimental schemes to implement single-photon added states using (a) beam-splitter and (b) NDPA. $\rho_{\rm in}$ is the input state, $\rho_{\rm single}$ the single-photon source, and $\rho_c$ the output state conditioned on the detection of (a) no-photon and (b) single-photon. $a_s$: signal mode, $a_i$: idler mode, BS: beam-splitter, PD: photo-detector.}
\label{fig:fig0}
\end{figure}

\subsection{beam-splitter scheme}
In Refs. [20,21], Dakna {\it et al.} proposed to use a beam-splitter together with $n$-photon Fock-state input so as to conditionally implement $n$-photon added scheme. 
Here, we restrict consideration to single-photon added scheme, as it seems practically feasible due to the single-photon sources available [28-30]. 
Suppose that a certain state $|\Psi\rangle$ is injected at one input and a single photon $|1\rangle$ at the other input to a beam splitter. [See Fig.1~(a).] 
If one performs photo-detection at one output of the beam splitter and detects no photons, then the input single photon must have traveled to the other output channel, 
which conditionally implements the single-photon addition to the input field as $a^\dag|\Psi\rangle$.

Now, there are certain realistic conditions that must be considered to discuss nonclassical properties of the conditional state in practice. 
First, the injected single photon source may be in a mixed state, particularly, a mixture of single photon and vacuum, $\rho_{\rm single}=p_S|1\rangle\langle1|+(1-p_S)|0\rangle\langle0|$, as shown in Refs. [28-30]. 
Secondly, a photon detector has the quantum efficiency $\eta$, thus no photo-detection collapses the system not to a vacuum state but to a mixture of number states, 
as represented by a POVM measurement, $\Pi^0=\sum_{n=0}^\infty(1-\eta)^n|n\rangle\langle n|=:e^{-\eta a^\dag a}:$.

The beam splitter action, $B_{12}$, can be represented by
\begin{eqnarray}
\begin{pmatrix}&b_1\\&b_2
\end{pmatrix}
=B_{12}^\dag \begin{pmatrix}&a_1\\&a_2
\end{pmatrix}B_{12}
=\begin{pmatrix}
&\cos\theta&\sin\theta\\&-\sin\theta&\cos\theta
\end{pmatrix}
\begin{pmatrix}&a_1\\&a_2
\end{pmatrix},
\end{eqnarray}
where $a_1$ and $a_2$ are input modes, $b_1$ and $b_2$ output modes, and $\cos\theta$ ($\sin\theta$) denote the transmissivity (reflectivity) of the beam splitter. 
If the initial input state is $\rho_{\rm in}\otimes\rho_{\rm single}$, the conditional output state becomes 
\begin{eqnarray}
\rho_c=\frac{1}{P_{\rm ND}}{\rm Tr}_2\{\Pi_2^0\cdot B_{12}\rho_{\rm in}\otimes\rho_{\rm single}B_{12}^\dag\},
\label{eqn:CS}
\end{eqnarray}
with the non-detection probability $P_{\rm ND}$ given by 
\begin{eqnarray}
P_{\rm ND}={\rm Tr}_{1,2}\{\Pi_2^0\cdot B_{12}\rho_{\rm in}\otimes\rho_{\rm single}B_{12}^\dag\}.
\label{eqn:NDP}
\end{eqnarray}

{\bf (i) Single-photon-added coherent states}\\
When the input state is a coherent state, $\rho_{\rm in}=|\alpha\rangle\langle\alpha|$, the conditional output state is obtained by a direct calculation as
\begin{eqnarray}
\rho_c=\frac{1}{N}\left[sa^\dag|\beta\rangle\langle\beta|a+f|\beta\rangle\langle\beta|+c|\beta\rangle\langle\beta|a+
ca^\dag|\beta\rangle\langle\beta|\right],\nonumber\\
\end{eqnarray}
where
\begin{eqnarray}
s&&\equiv Rp_S\nonumber\\
f&&\equiv p_S(1-\eta)T\left(1+(1-\eta)R\alpha^2\right)+(1-p_S)\nonumber\\
c&&\equiv-Rp_S(1-\eta)\beta\nonumber\\
N&&\equiv p_S\left(1-T\eta+RT\eta^2\alpha^2\right)+1-p_S,
\end{eqnarray}
with $\beta=\alpha\cos\theta, T\equiv \cos^2\theta$ (transmittance), and $R\equiv \sin^2\theta$ (reflectance). 
The non-detection probability turns out to be
\begin{eqnarray}
P_{\rm ND}=Ne^{-R\eta\alpha^2}.
\end{eqnarray}
To investigate the behaviors of two nonclassical properties, $Q_1^m$ and $Q_2^m$ in Eqs.~(\ref{eqn:ONCD1}) and~(\ref{eqn:NCD2}), 
we just need to calculate  $\langle a^{m}\rangle$ and $\langle a^{\dag m}a^{m}\rangle$, 
which are obtained as 
\begin{eqnarray}
\langle a^{m}\rangle=\frac{1}{N}\left[s\beta^m(\beta^2+m+1)+f\beta^m+c\beta^{m-1}(2\beta^2+m)\right],\nonumber\\
\end{eqnarray}
and
\begin{eqnarray}
\langle a^{\dag m}a^{m}\rangle=\frac{1}{N}\left[s\beta^{2m-2}((\beta^2+m)^2+\beta^2)+f\beta^{2m}\right.\nonumber\\ +2c\beta^{2m-1}(\beta^2+m)\left.\right].
\end{eqnarray}
The anti-normal ordered moment $\langle a^{m} a^{\dag m}\rangle$ in Eq.~(\ref{eqn:ONCD1}) can be evaluated in terms of 
the normal ordered ones using the relation 
\begin{eqnarray}
a^{m}a^{\dag m}=\sum_{p=0}^{m}\frac{(m!)^2}{(m-p)!(p!)^2}a^{\dag p} a^p.
\label{eqn:ANM}
\end{eqnarray}

\begin{figure}
\includegraphics[width=3.0in,keepaspectratio=true]{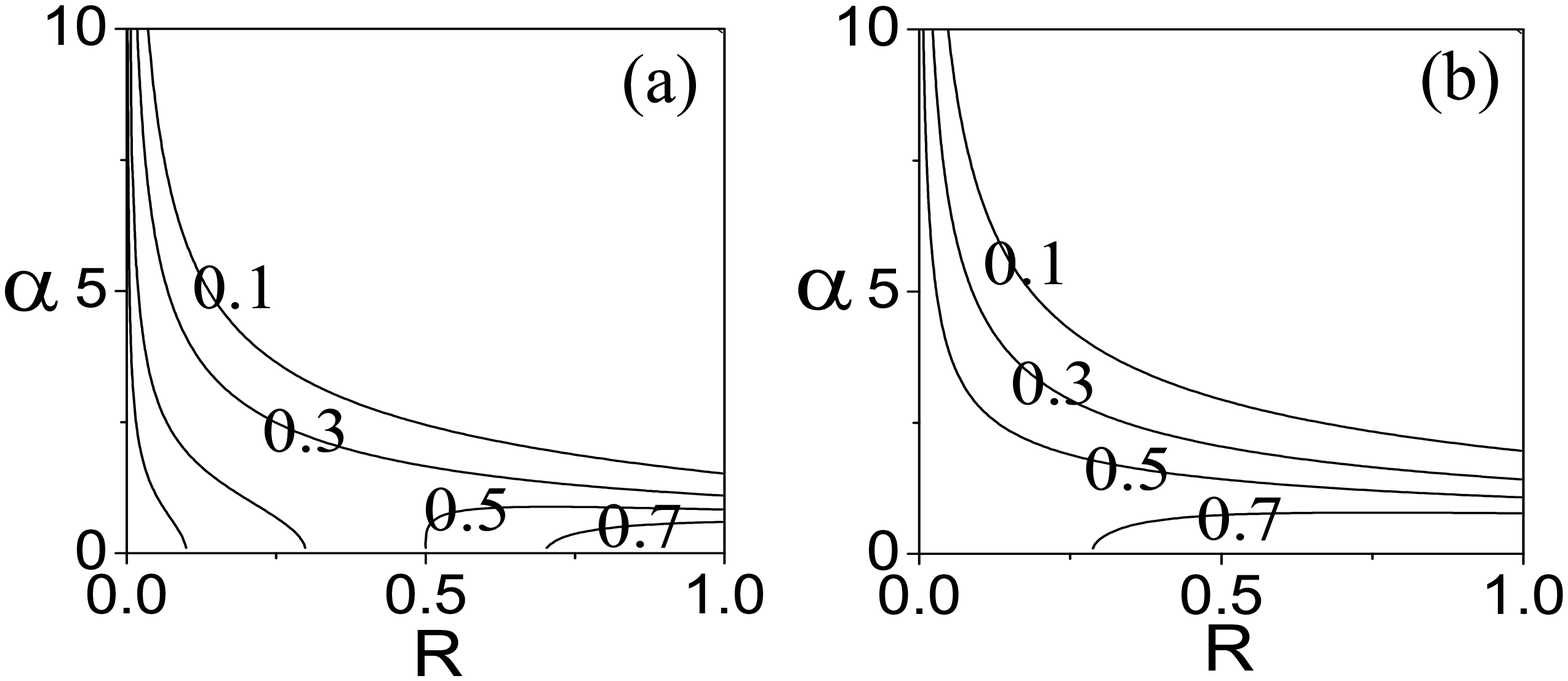}
\caption{Contour plot for the probability $P_{\rm ND}$ to generate an approximate SACS conditioned on the no-detection of photons as a function of $\alpha$ and $R=\sin^2\theta$ (reflectance of beam splitter) for (a) $\eta=1$ and $p_S=1$, and (b) for $\eta=0.6$ and $p_S=0.7$.}
\label{fig:fig1}
\end{figure}

\begin{figure}
\includegraphics[width=3.0in,keepaspectratio=true]{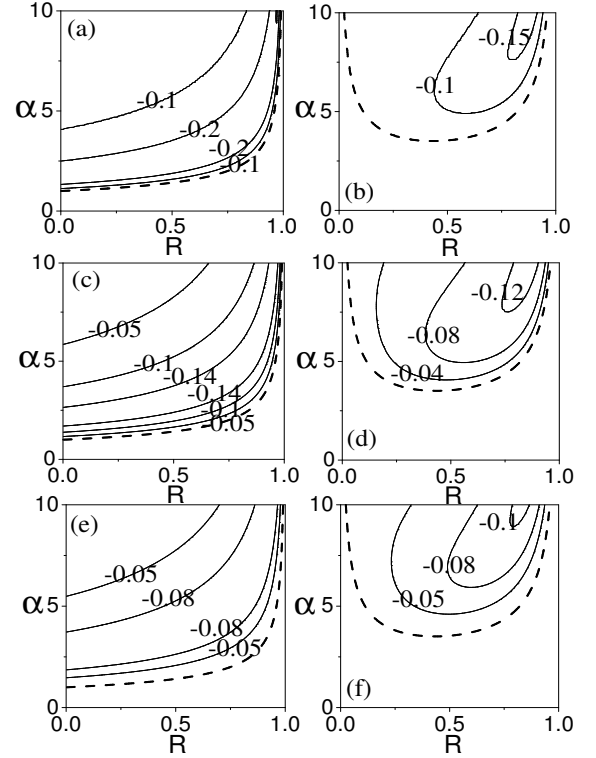}
%\centerline{\scalebox{1.4}{\includegraphics{fig3_1.eps}}}
\caption{Contour plot of $Q_1^m$ of SACS as a function of $\alpha$ and $R=\sin^2\theta$ for (a),(b) $m=1$ (quadrature squeezing), (c),(d) $m=2$ (Hillery's amplitude squared squeezing), and (e),(f) $m=3$. The left-column plots [(a), (c), and (e)] are for the ideal case of $\eta=1$ and $p_S=1$, and the right-column ones [(b), (d), and (f)] for $\eta=0.6$ and $p_S=0.7$. The dashed lines ($Q_1^m=0$) represent the boundary between classical and nonclassical regimes.}
\label{fig:fig2}
\end{figure}

In particular, we compare the ideal case of $\eta=1$ and $p_S=1$ with the nonideal case of $\eta=0.6$ and $p_S=0.7$ throughout this paper. 
The cases with other values of $\eta$ and $p_S$ can be readily inferred by the comparison. 
Note that the single-photon probability, $p_S$, was reported to be up to $p_S=0.69$ in Ref. [29]. 

In Fig.~2, the contour plot of the conditional probability $P_{\rm ND}$ is displayed as a function of $\alpha$ and $R$ (reflectance). 
When the input amplitude $\alpha$ is large, the non-detection probability, $P_{\rm ND}$, becomes smaller with the higher reflectance $R$. 
On the other hand, when $\alpha$ is small, $P_{\rm ND}$ becomes larger with the higher $R$. 
This behavior can be understood by the fact that the coherent-state input has the mean photon number ${\bar n}=|\alpha|^2$ and the other non-ideal single photon source 
${\bar n}=p_S$:  
For $|\alpha|^2>p_S$, a larger number of photons impinges on the detector with higher $R$, thereby reducing $P_{\rm ND}$, and vice versa.

In Fig.~3, the contour plot of $Q_1^m$ is displayed for $m=1$ (quadrature squeezing), $m=2$ (Hillery's amplitude squared squeezing), and $m=3$ as examples. 
In the ideal case of $\eta=1$ and $p_S=1$, the negativity of $Q_1^m$ becomes largest along the lines where $\alpha$ is monotonically increasing with $R$, 
and this tendency remains the same regardless of the order $m$. 
In the non-ideal case of $\eta<1$ and $p_S<1$, the parameter space of $\alpha$ and $R$, where negative values appear above the dashed lines ($Q_1^m=0$), 
is narrowed. However, we see that the negativity of $Q_1^m$ is still substantial at the realistic values of $\eta=0.6$ and $p_S=0.7$. 
One practical disadvantage, though, is that in the non-ideal cases, the optimal negativity appears in the region where the conditional probability $P_{\rm ND}$ is very small.
[Cf. Fig.~2.] 
Nevertheless, if one is ready to slightly sacrifice the degree of negativity in $Q_1^m$, 
the phase-sensitive nonclassical effects seem observable in high orders. 
For instance, with $\alpha=5$ and $R=0.5$, we obtain $Q_1^{m=1}\approx-0.098$, $Q_1^{m=2}\approx-0.08$, and $Q_1^{m=3}\approx-0.062$ 
at the conditional probability $P_{\rm ND}\approx1.3\times10^{-3}$.

\begin{figure}
\includegraphics[width=3.0in,keepaspectratio=true]{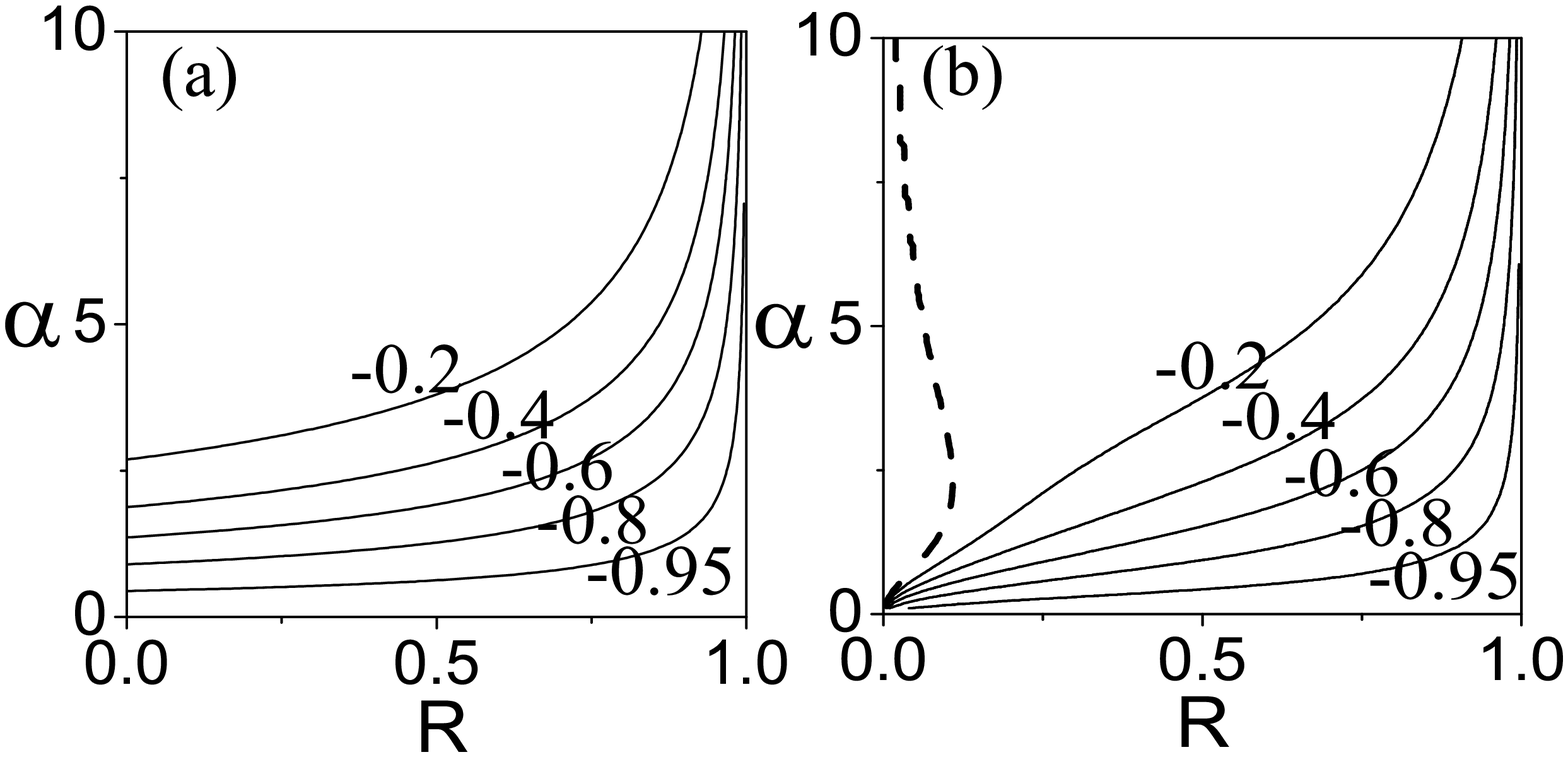}
\caption{Contour plot of $Q_2^{m=4}$ of SACS as a function of $\alpha$ and $R=\sin^2\theta$ for (a) $\eta=1$ and $p_S=1$, and (b) $\eta=0.6$ and $p_S=0.7$. 
The dashed lines ($Q_2^m=0$) represent the boundary between classical and nonclassical regimes.}
\label{fig:fig3}
\end{figure}

In Fig.~4, we show the contour plot of the phase-insensitive nonclassical effect, $Q_2^m$, for $m=4$ as an example. 
$Q_2^m$ shows similar behaviors in all the other orders of $m$.
It turns out that the nonclassical effects, $Q_2^m$, are less sensitive to the practical imperfections than $Q_1^m$. 
Moreover, the condition for optimal negativity of $Q_2^m$ is more favorable in view of the conditional probability $P_{\rm ND}$. [Cf. Fig.~2.] 
For instance, even at much lower values of $\eta=0.3$ and $p_S=0.3$, with $\alpha=1.5$ and $R=0.8$, we find 
$Q_2^{m=2,3,4,5}\approx-0.14,-0.46,-0.65,-0.76$ 
at the conditional probability $P_{\rm ND}\approx0.58$. 
Generally, the degree of negativity in $Q_2^m$ becomes larger with the order $m$ unlike $Q_1^m$.

\begin{figure}
\includegraphics[width=3.0in,keepaspectratio=true]{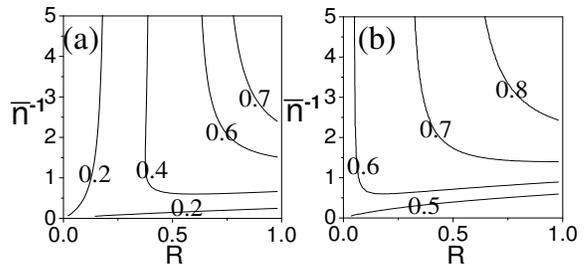}
\caption{The probability $P_{\rm ND}$ to generate an approximate SATS conditioned on the no-detection of photons as a function of $1/\bar n$ and $R=\sin^2\theta$ for (a) $\eta=1$ and $p_S=1$, and (b) $\eta=0.6$ and $p_S=0.7$.}
\label{fig:fig4}
\end{figure}

{\bf (ii) Single-photon-added thermal states}\\
When the input state to the beam splitter is a thermal state, $\rho_{\rm in}=(1-e^{-\beta})e^{-\beta a^\dag a}$,  with the average photon number $\bar{n}=\frac{1}{e^\beta-1}$, we obtain the $m$-photon coincidence rate at the output, using Eqs.~(\ref{eqn:CS}) and~(\ref{eqn:NDP}), as 
\begin{eqnarray}
&&\langle a_1^{\dag m}a_1^m\rangle={\rm Tr}_1\{a_1^{\dag m}a_1^m\rho_c\}\nonumber\\
&&=\frac{{\rm Tr}_{1,2}\{a_1^{\dag m}a_1^m\Pi_2^0\cdot B_{12}\rho_{\rm in}\otimes\rho_{\rm single}B_{12}^\dag\}}{{\rm Tr}_{1,2}\{\Pi_2^0\cdot B_{12}\rho_{\rm in}\otimes\rho_{\rm single}B_{12}^\dag\}}\equiv\frac{D_m}{D_0},\nonumber\\
\end{eqnarray}
where
\begin{eqnarray}
D_m
=\frac{m!T^{m-1}{\bar n}^{-1}}{({\bar n}^{-1}+\eta R)^{m+2}}\left[T({\bar n}^{-1}+\eta R)(1-\eta p_ST+2m\eta p_S R)\right.\nonumber\\+mp_SR({\bar n}^{-1}+\eta R)^2+(m+1)\eta^2p_SRT^2\left.\right],\nonumber\\
\end{eqnarray}
and the conditional probability is given by $P_{\rm ND}=D_0$.

In Fig.~5, we show the contour plot of the conditional probability $P_{\rm ND}$ as a function of ${\bar n}^{-1}$ and $R$. 
As is the case with the coherent state input [Cf. Fig.~2], the conditional probability becomes larger with lower $R$ if the mean photon number ${\bar n}$ is large, and vice versa. In Fig.~6, we show the contour plot of $Q_2^m$ for $m=1$ (sub-Poissonian), $m=2$, and $m=3$ as examples. 
As the order $m$ is increased, the parameter space for $Q_2^m<0$ is narrowed, as explained in Sec.~III B, but the nonclassical effects appear in a sufficiently broad range of parameters even in non-ideal conditions. Generally, the range of reflectance $R$ in which the negativity appears becomes wider for a lower value of mean photon number ${\bar n}$.

\begin{figure}
\includegraphics[width=3.0in,keepaspectratio=true]{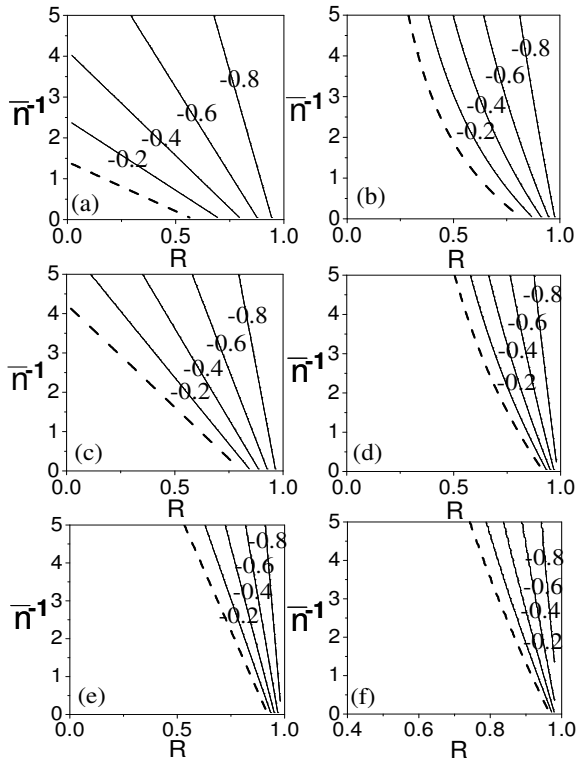}
\caption{Contour plot of $Q_2^{m}$ of SATS as a function of $1/\bar n$ and $R=\sin^2\theta$ for (a),(b) $m=1$ (sub-Poissonian statistics), (c),(d) $m=2$ , and (e),(f) $m=3$. The left-column plots [(a), (c), and (e)] are for the ideal case of $\eta=1$ and $p_S=1$, and the right-column ones [(b), (d), and (f)] for $\eta=0.6$ and $p_S=0.7$.}
\label{fig:fig5}
\end{figure}

\subsection{ NDPA scheme}
Secondly, let us briefly consider the case that the non-degenerate parametric amplifier (NDPA) is used to implement single-photon added scheme, as demonstrated in Refs. [23-26]. The NDPA is described by the Hamiltonian $H=i\hbar\xi\left(a_s^\dag a_i^\dag-a_sa_i\right)$, where $\xi$ denotes the classical pumping strength, and $a_s$ and $a_i$ are the signal and the idler modes at the NDPA. [See Fig.1~(b).]
When the pumping is sufficiently low, the input field $|\Psi\rangle_s$ in the signal and the vacuum $|0\rangle_i$ in the idler are converted to an entangled state  $|\Psi(t)\rangle=e^{-iHt/\hbar}|\Psi\rangle_s|0\rangle_i\approx\left[1+\xi t\left(a_s^\dag a_i^\dag-a_sa_i\right)\right]|\Psi\rangle_s|0\rangle_i$. 
Then, conditioned on a single-photon detection in the idler mode, the signal mode collapses to $\langle1_i|\Psi(t)\rangle\approx\xi t a_s^\dag|\Psi\rangle_s$, 
which implements the single-photon-addition to the input state $|\Psi\rangle_s$.

In a realistic situation, the nonideal detection such as the dark counts and multi-photon counts in the idler mode deforms the target state $a_s^\dag|\Psi\rangle_s$ to a certain mixed state. In Refs. [23-26], Zavatta {\it et al.} particularly considered the limited efficiency,  not only in the preparation of the target state but also in the homodyne detection of the generated state, by a beam-splitter model: the actually generated state is equivalent to the output from a beam splitter of transmissivity $\eta$ to which the ideal target state $a_s^\dag|\Psi\rangle_s$ and the vacuum state $|0\rangle_s$ are injected.
Here, $\eta$ denotes the overall, empirical, efficiency of the experimental scheme. This modeling showed an excellent agreement with the experimental data [23-26], so we adopt it here to investigate the nonclassical properties of the single-photon-added classical states.

\begin{figure}
\includegraphics[width=3.0in,keepaspectratio=true]{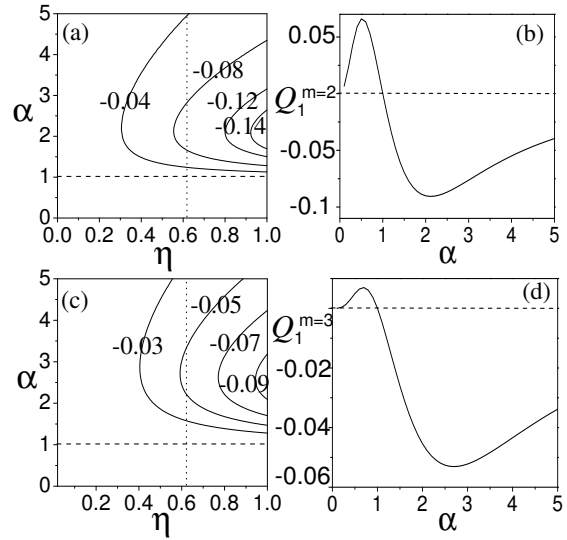}
\caption{Plot of $Q_1^m$ as a function of $\alpha$ and $\eta$ for (a),(b) $m=2$ (Hillery's amplitude squared squeezing), and (c),(d) $m=3$ in the NDPA scheme. The dashed lines represent the boundary ($Q_1^m=0$) between classical and nonclassical regimes. In (a) and (c), the dotted line corresponds to $\eta=0.62$ for which $Q_1^m$ is plotted again as a function of $\alpha$ in (b) and (d), respectively. }
\label{fig:fig6}
\end{figure}

The above-mentioned BS model yields the relation $a_D=\sqrt{\eta}a+\sqrt{1-\eta}v$, where $a_D$ is the actually generated mode, $a$ the target mode, and $v$ the vacuum mode. 
Now, it is straightforward to obtain the relations 
$\langle a_D^{m}\rangle=\sqrt{\eta}^m\langle a^{m}\rangle$ and $\langle a_D^{\dag m}a_D^{m}\rangle={\eta}^m\langle a^{\dag m}a^{m}\rangle$.

{\bf (i) Case of $Q_1^m$}\\
For the phase-sensitive effects, $Q_1^m$, therefore, the limited efficiency $\eta$ does not affect the region where the negativity emerges in comparison to the pure-state cases: 
${\eta}^m$ becomes an overall factor in the numerator of $Q_1^m$ from Eq.~(\ref{eqn:ONCD1}). 
This means that even in the presence of nonideal efficiency, for the coherent state input,
$Q_1^m$ becomes negative in all orders of $m$ for $\alpha>1$ regardless of $\eta$. 
However, due to the  anti-normal moment $\langle a_D^{\dag m}a_D^{m}\rangle$, 
which contains the terms proportional to $\eta^p$ in the range of $p=0,\dots,m$ [see Eq.~(\ref{eqn:ANM})], 
the degree of negativity is affected by the efficiency $\eta$. 
In Fig.~7, we show the contour plot of $Q_1^m$ as a function of $\alpha$ and $\eta$ for $m=2$ (Hillery's amplitude squared squeezing), and $m=3$ as examples. 
Apparently, the degree of negativity is reduced by the nonideal efficiency $\eta$, 
however, it is still substantial to observe. 
In Figs.~7 (b) and (d), we again plot $Q_1^m$ as a function of $\alpha$ particularly at the efficiency $\eta=0.62$ that was reported in Ref. [25]. 
Those plots clearly demonstrate the higher-order nonclassical properties of the states generated by Zavatta {\it et al.}. 

{\bf (ii) Case of $Q_2^m$}\\
For the phase-insensitive effects, $Q_2^m$, on the other hand, 
the limited efficiency $\eta$ does not play any role: the same overall factors ${\eta}^m$ from the numerator and the denominator in Eq.~(\ref{eqn:NCD2}) cancel out. 
Therefore, the degree of nonclassicality does not change, and the analysis given to the pure state cases (coherent and thermal) in Sec.~III is valid regardless of $\eta$.

\section{Conclusions}
In this paper, we have investigated two generalized classes of nonclassical properties, 
one phase-sensitive and the other phase-insensitive, for the single-photon added classical (coherent and thermal) states. 
We have shown that all higher-order nonclassical effects can be observed using SACSs and SATSs. In particular, we analyzed two feasible experimental schemes, beam-splitter scheme and NDPA scheme, and demonstrated the robustness of those nonclassical effects even in the presence of experimental imperfections. 
Our analysis also showed that the experimental results reported by Zavatta {\it et al.} already imply the high-order nonclassical effects [23-26].

To experimentally observe the nonclassical properties studied here, 
one needs to measure two different types of moments, $\langle a^{m}\rangle$ and $\langle a^{\dag m}a^{m}\rangle$, as can be seen from Eqs.~(\ref{eqn:ONCD1}) and~(\ref{eqn:NCD2}). Recently, Shchukin and Vogel proposed the balanced homodyne correlation measurement 
to efficiently measure general normally-ordered moments [31], which could be employed to measure those moments involved. 
For the case of $\langle a^{\dag m}a^{m}\rangle$, one can simply use the $m$-photon coincidence measurement after dividing the signal mode to $m$-outputs via an array of beam-splitters. 
Then, the measured $Q_2^m$, as defined in Eq.~(\ref{eqn:NCD2}), becomes insensitive to the quantum efficiency of photo detectors, which is a practical advantage. 
However, it is always desirable to have a higher efficiency in order to enhance the counting rates.

\section*{Acknowledgments}
JL and JK were partially supported by the IT R\&D program of MKE/IITA
(2008-F-035-02). 
HN is supported by an NPRP grant from the Qatar National Research Fund.  

*email: hyunchul.nha@qatar.tamu.edu

\end{document}